\documentclass[aip,apl,reprint]{revtex4-1}

\usepackage{graphicx}
\usepackage{amsmath}
\usepackage{textcomp} 
\usepackage{xcolor}
\usepackage{ulem}

\begin{document}

\title{Heat-induced damping modification in YIG/Pt hetero-structures}

\author{M.~B.~Jungfleisch}
\email{jungfleisch@physik.uni-kl.de}
\affiliation{Fachbereich Physik and Landesforschungszentrum OPTIMAS, Technische Universit\"at Kaiserslautern, 67663
Kaiserslautern, Germany}

\author{T.~An}
\affiliation{Institute for Materials Research, Tohoku University, Sendai 980-8577, Japan}
\affiliation{CREST, Japan Science and Technology Agency, Tokyo 102-0076, Japan}

\author{K.~Ando}
\affiliation{Institute for Materials Research, Tohoku University, Sendai 980-8577, Japan}

\author{Y.~Kajiwara}
\affiliation{Institute for Materials Research, Tohoku University, Sendai 980-8577, Japan}
\affiliation{CREST, Japan Science and Technology Agency, Tokyo 102-0076, Japan}

\author{K.~Uchida}
\affiliation{Institute for Materials Research, Tohoku University, Sendai 980-8577, Japan}
\affiliation{PRESTO, Japan Science and Technology Agency, Saitama 332-0012, Japan}

\author{V.~I.~Vasyuchka}
\affiliation{Fachbereich Physik and Landesforschungszentrum OPTIMAS, Technische Universit\"at
Kaiserslautern, 67663 Kaiserslautern, Germany}

\author{A.~V.~Chumak}
\affiliation{Fachbereich Physik and Landesforschungszentrum OPTIMAS, Technische Universit\"at
Kaiserslautern, 67663 Kaiserslautern, Germany}

\author{A.~A.~Serga}
\affiliation{Fachbereich Physik and Landesforschungszentrum OPTIMAS, Technische Universit\"at
Kaiserslautern, 67663 Kaiserslautern, Germany}

\author{E.~Saitoh}
\affiliation{Institute for Materials Research, Tohoku University, Sendai 980-8577, Japan}
\affiliation{CREST, Japan Science and Technology Agency, Tokyo 102-0076, Japan}
\affiliation{WPI Advanced Institute for Materials Research, Tohoku University, Sendai 980-8577, Japan}

\affiliation{Advanced Science Research Center, Japan Atomic Energy Agency, Tokai 319-1195, Japan}

\author{B.~Hillebrands}
\affiliation{Fachbereich Physik and Landesforschungszentrum OPTIMAS, Technische Universit\"at
Kaiserslautern, 67663 Kaiserslautern, Germany}

\date{\today}

\begin{abstract}

We experimentally demonstrate the manipulation of magnetization relaxation utilizing a temperature difference across the thickness of an yttrium iron garnet/platinum (YIG/Pt) hetero-structure: the damping is either increased or decreased depending on the sign of the temperature gradient. This effect might be explained by a thermally-induced spin torque on the magnetization precession. The heat-induced variation of the damping is detected by microwave techniques as well as by a DC voltage caused by spin pumping into the adjacent Pt layer and the subsequent conversion into a charge current by the inverse spin Hall effect.
\end{abstract}

\maketitle

Due to their interesting underlying physics and potential applications in magnon spintronics the spin Hall effect (SHE) and the inverse spin Hall effect (ISHE) attracted considerable attention in the last years. \cite{ Saitoh-2006, Sandweg2} Magnon spintronics is a new, emerging field in spintronics, that utilizes magnons (quanta of spin waves) as carriers of angular momentum. The combination of spin pumping and inverse spin Hall effect turned out to be a well suited technique for the detection of magnons beyond the wavenumber limitations of most other methods. \cite{Sandweg2} The recent discovery of the spin Seebeck effect (SSE) in magnetic insulators demonstrates the importance of heat currents in spintronics and opened up the new field of spin caloritronics. \cite{Uchida2010, Uchida_long}

A key objective in the field of magnon spintronics is the control of magnetization relaxation and generation of spin waves. To compensate spin-wave damping a common method is parametric amplification. \cite{Melkov, Sandweg2} 
Recently, it was reported, that propagating spin waves can also be amplified by injecting a spin current  due to the SHE and the spin-transfer torque (STT) effect \cite{MWu2011} and by the SSE. \cite{Rezende_PRL_2011} Spin relaxation was manipulated by SHE and STT in Ni$_{81}$Fe$_{19}$ \cite{Ando_PRL_2008} and by thermally-induced interfacial spin transfer in yttrium iron garnet/platinum (YIG/Pt) structures. \cite{Wu2012} In all these experiments, magnetization dynamics is measured by using microwave techniques. However, two-magnon scattering can lead to the excitation of secondary spin waves with much higher wavevectors as it has been shown in Refs.~\cite{Jungfleisch, Chumak}. Even though these secondary waves contribute significantly to spin pumping, \cite{Tserkovnyak, Costache} they cannot be detected by inductive antennas. Therefore, microwave measurements do not necessarily give a thorough insight into magnetization dynamics.

\begin{figure}[b]
\includegraphics[width=0.8\columnwidth]{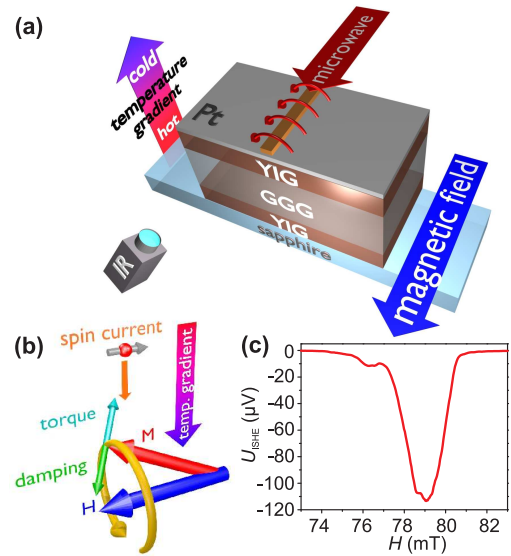}
\caption{\label{fig:setup} (Color online) (a) Schematic illustration of the experimental setup. (b) Possible mechanism for the heat-induced damping variation in the YIG film. \textbf{M} denotes the magnetization. (c) Typical example for a measurement of $U_\mathrm{ISHE}$ as a function of the applied magnetic field ${H}$.} 
\end{figure}

In this Letter, we report  on the thermal manipulation of spin-wave relaxation measured by both microwave techniques as well as by spin pumping. The investigated sample consists of a YIG/GGG/YIG/Pt hetero-structure. A temperature difference applied across the thickness of this structure leads to the longitudinal SSE: an imbalance between the magnon and electron temperatures causes a spin current across the YIG/Pt interface. \cite{Uchida2010, Uchida_long} The generated spin current transfers angular momentum and, consequently,  they might exert a torque  on the magnetization (see Fig.~\ref{fig:setup}(b)). As a result, the magnetization precession can either be enhanced or suppressed depending on the sign of the temperature difference and, thus, the direction of the spin current. This change in the damping is equivalent to a variation of the ferromagnetic resonance linewidth $\Delta H_\mathrm{FMR}$ that is measured by microwave reflection as well as by spin pumping.

\begin{figure}[t]
\includegraphics[width=1\columnwidth]{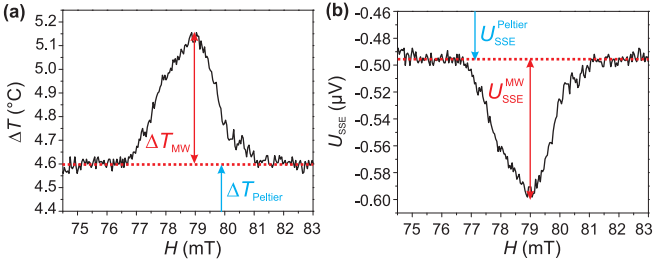}
\caption{\label{fig:sub} (Color online) (a) Typical example for the temperature difference $\Delta T$ across the sample as a function of the applied magnetic field $H$. $\Delta T_\mathrm{Peltier}$ denotes the temperature difference applied by the Peltier element. At the ferromagnetic resonance field around 79~mT the magnetization precession causes additional heating denoted as $\Delta T_\mathrm{MW}$. (b) Calculated spin Seebeck voltage $U_\mathrm{SSE}$ composed of $U_\mathrm{SSE}^\mathrm{Peltier}$ generated by $\Delta T_\mathrm{Peltier}$ and  $U_\mathrm{SSE}^\mathrm{MW}$ created by $\Delta T_\mathrm{MW}$.} 
\end{figure}

A sketch of the experimental setup is shown in Fig.~\ref{fig:setup}(a). A 2.1~$\mu$m thick YIG film was grown by means of liquid phase epitaxy on both sides of a 500~$\mu$m thick gadolinium gallium garnet (GGG) substrate. Using molecular beam epitaxy, we then deposited a Pt layer of 10~nm thickness on one side of the sample, fully covering one of the YIG surfaces. As shown in Ref.~\cite{PRL_Chien_2012}  the Pt layer might show ferromagnetic behavior on ferromagnetic insulators due to magnetic proximity effects. 
However, as shown in Ref.~\cite{Kikkawa}, a possible contamination by the anomalous Nernst effect is negligibly small compared to the longitudinal SSE contribution. A Peltier element, that is mounted on top of the Pt layer, generates a temperature difference across the multilayer. In order to enhance the temperature flow from the sample, the second YIG surface is covered with a 
sapphire substrate that is connected to a heat bath (sapphire is a good thermal conductor). 
The second YIG layer neither influences the magnetic nor the electric measurements but it should be noted that the temperature difference is applied across the entire Pt/YIG/GGG/YIG sample stack. The magnetization precession is excited by a 500~$\mu$m wide copper microstrip antenna that is placed above the Pt layer with an intervening 
isolation layer (see Fig.~\ref{fig:setup}(a)). The temperature difference is monitored using an infrared camera, calibrated by two thermocouples.

The experiment is performed as follows: an external magnetic field $\textbf{H}$ is applied perpendicularly to the YIG waveguide in the YIG film plane. The magnetization precession is driven by the alternating magnetic field  $\textbf{h}(t)$ of a continuous microwave signal (see Fig.~\ref{fig:setup}(a) and (b)) with a carrier frequency of 4~GHz and powers of $P_\mathrm{MW}=+14$~dBm, +20~dBm, and +25~dBm. While sweeping the external magnetic field $H$, a temperature difference across the sample thickness is applied and recorded by the infrared camera. The electric voltage due to the ISHE, $U_\mathrm{ISHE}$, and the microwave reflection are measured simultaneously.

\begin{figure}[b]

\includegraphics[width=1\columnwidth]{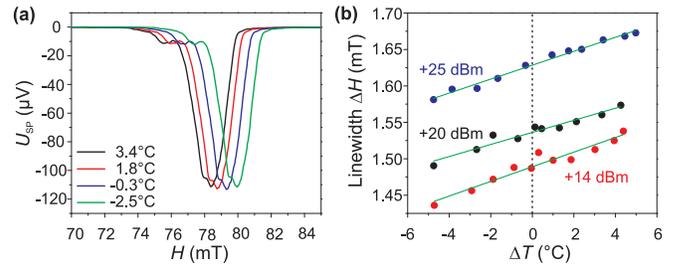}

\caption{\label{fig:results} (Color online) (a) Typical spectra for 4 different temperature differences at +25 dBm. For positive (negative) temperature differences, the Pt is colder (hotter) than the YIG.
(b) Measured resonance linewidth $\Delta H$ as a function of the temperature difference $\Delta T$ for different microwave powers. }
\end{figure}

In the experiment, two different mechanisms contribute to the spin current: the spin Seebeck effect and the spin pumping effect. The SSE originates from the difference between the effective magnon temperature $T_\mathrm{m}$  and the effective electron temperature $T_\mathrm{e}$ at the YIG/Pt interface.  \cite{Uchida_long, Uchida2010, Uchida_JAP}  This temperature difference $\Delta T$ is created in two different ways in our experiment (see Fig.~\ref{fig:sub}(a)): by the Peltier element (denoted by $\Delta T_\mathrm{Peltier}$ in the following) and by heating due to magnetization precession in resonance condition of the YIG film at $H_\mathrm{FMR}$ (denoted by $\Delta T_\mathrm{MW}$). The second mechanism to create a spin current is spin pumping by the externally excited coherent magnetization precession. \cite{Tserkovnyak} Irrespective of its origin, the net spin current $\textbf{J}_\mathrm{s}$ injected into the Pt layer is transformed into a conventional charge current $\textbf{J}_\mathrm{c}$, perpendicular to both $\textbf{J}_\mathrm{s}$ and $\textbf{H}$, by the ISHE (see Fig~\ref{fig:setup}(c)). \cite{Hirsch, Jungfleisch} As a result, charges accumulate at the edges of the Pt layer and a voltage $U_\mathrm{ISHE}=U_\mathrm{SP}+U_\mathrm{SSE}$, composed of a spin pumping contribution $U_\mathrm{SP}$ and a SSE contribution $U_\mathrm{SSE}$, can be measured (Fig.~\ref{fig:setup}(c)). The voltage  $U_\mathrm{SSE}=U_\mathrm{SSE}^\mathrm{MW}+U_\mathrm{SSE}^\mathrm{Peltier}$ itself consists of the voltage $U_\mathrm{SSE}^\mathrm{MW}$ generated by heating $\Delta T_\mathrm{MW}$ due to magnetization precession in resonance and $U_\mathrm{SSE}^\mathrm{Peltier}$ generated by the temperature difference $\Delta T_\mathrm{Peltier}$. 
In order to distinguish between the different contributions to $U_\mathrm{ISHE}=U_\mathrm{SSE}^\mathrm{Peltier}+U_\mathrm{SSE}^\mathrm{MW}+U_\mathrm{SP}$ the following procedure has been used: for each temperature difference created with the Peltier element, $U_\mathrm{ISHE}$ and $\Delta T$ were recorded. From the off-resonance condition ($H > H_\mathrm{FMR}$), we can deduce a linear relation between $U_\mathrm{SSE}=U_\mathrm{SSE}^\mathrm{Peltier}$ and $\Delta T$. Using this linear SSE relation \cite{Uchida_long, Uchida2010, Uchida_JAP} enables us to recalculate the corresponding voltage $U_\mathrm{SSE}^\mathrm{MW}$ at $H_\mathrm{FMR}$ and, thus, the spin pumping voltage $U_\mathrm{SP}$, respectively. 
In Fig.~\ref{fig:sub} the evolution of the temperature difference as a function of the applied magnetic field $H$ and the corresponding SSE voltage $U_\mathrm{SSE}$ are shown for a microwave power of $P_\mathrm{MW}=+25$~dBm. Figure~\ref{fig:sub}(a) clearly shows, that, in addition to the applied temperature difference of $\Delta T_\mathrm{Peltier}\approx 4.6^\circ$C, the temperature rises at $H_\mathrm{FMR}$ by an additional value of $\Delta T_\mathrm{MW}\approx 0.6^\circ$C. The corresponding voltages $U_\mathrm{SSE}^\mathrm{Peltier}$ and $U_\mathrm{SSE}^\mathrm{MW}$ are illustrated in Fig.~\ref{fig:sub}(b). For $\Delta T = 0$, the FMR driven spin pumping contribution $U_\mathrm{SP}$ is dominant ($U_\mathrm{SP}/(U_\mathrm{SP}+U_\mathrm{SSE}) \approx 99.9 \%$).

The heat-induced spin current affects our measurements in two different ways. On one side, it generates a voltage $U_\mathrm{SSE}$ independent of the absolute value of the externally applied magnetic field $\textbf{H}$ (crossing zero field results in a change of the polarity of $U_\mathrm{SSE}$ according to the SSE \cite{Uchida_long, Uchida2010, Uchida_JAP}), and on the other side, it most likely exerts a torque on the magnetization, resulting in the manipulation of the relaxation damping (see Fig.~\ref{fig:setup}(b)).

Figure~\ref{fig:setup}(c) shows a typical example for the measured ISHE voltage $U_\mathrm{ISHE}$ without externally applied  temperature difference. The voltage reaches its maximal absolute value at $H_\mathrm{FMR} \approx 79$~mT.  In Fig.~\ref{fig:results}(a) the recalculated $U_\mathrm{SP}$ data is shown as a function of the external magnetic field $H$ for four different measured temperature differences $\Delta T$. Heating and cooling the sample gives rise to a change of the saturation magnetization $M_\mathrm{S}$ resulting in a resonance peak shift to higher or lower magnetic fields.

As it is obvious from Fig.~\ref{fig:results}(a), not only one but several modes contribute to the spin pumping voltage $U_\mathrm{SP}$.  
Therefore, the envelope of $U_\mathrm{SP}$ is fitted for each temperature difference by a Gaussian function $f(x) = a\cdot \mathrm{exp}(-(x-b)^2 / (2c^2))$, where $c$ defines the linewidth $\Delta H$ which is a measure for 
the damping $\alpha$. The linewidth $\Delta H$ that is determined in this way, does not necessarily coincide with the real ferromagnetic resonance linewidth $\Delta H_\mathrm{FMR}$ but is proportional to it, i. e. $\Delta H \propto \Delta H_\mathrm{FMR}$. The linewidth $\Delta H$ as a function of the temperature difference $\Delta T$ is shown in Fig.~\ref{fig:results}(b) for different microwave powers $P_\mathrm{MW}$. It is clearly visible that the total linewidth $\Delta H$ decreases for one polarity of $\Delta T$ and increases for the other. We also analyzed each mode separately and we found that the qualitative behavior for each mode is the same.

As it is visible from Fig.~\ref{fig:results}(b), the variation of the linewidth $\Delta H$ per $1^\circ$C temperature difference (slope in Fig.~\ref{fig:results}(b)) is approximately the same for all microwave powers. For a temperature difference of $\Delta T \approx \pm 4^\circ$C, the linewidth changes about $6\%$, independent of the microwave power. This independency is expected since the generated spin currents are thermally induced and, thus, do not depend on the applied microwave power. \cite{Uchida2010} 
The damping, i. e., the linewidth at $\Delta T = 0$ is larger for higher microwave powers which is attributed to the onset of non-linear effects. \cite{Mills, Hu, Demidov_PRB}

\begin{figure}[t]
\includegraphics[width=0.60\columnwidth]{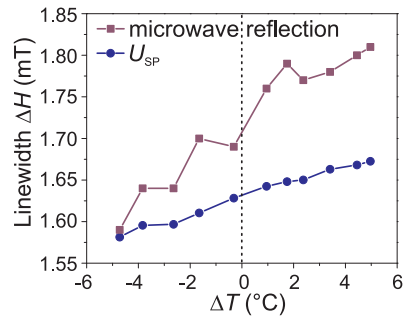}
\caption{\label{fig:mw} (Color online) Measured linewidth $\Delta H$ as a function of the measured temperature difference $\Delta T$ for reflected microwave signal and spin pumping voltage  $U_\mathrm{SP}$.}
\end{figure}

Now we compare how the linewidth alters under the influence of a longitudinal temperature difference measured by both spin pumping as well as microwave techniques. Since spin pumping is not sensitive to the spin-wave wavelength, the directly excited spin-wave modes as well as short-wavelength secondary waves contribute to the detected signal. \cite{Jungfleisch, Chumak} However, microwave reflection mainly detects the primary excited uniform mode. The results are summarized in Fig.~\ref{fig:mw}. For both measurement techniques, the linewidth qualitatively behaves the same in the investigated range of temperature differences. 
Nevertheless, one can see that the slopes of the two curves diverge leading to the assumption that the uniform FMR mode is mostly effected by the heat-induced damping modification. However, a quantitative statement is not possible. 

Assuming, that the observed heat-induced damping variation is due to thermal spin currents generated by the SSE, we calculate the variation of the magnetization relaxation in YIG/Pt hetero-structures based on the model developed in Ref.~\cite{Ando_PRL_2008}. We modify this model by substituting a heat-induced spin current for a SHE-generated spin current: the charge current $J_\mathrm{c}$ is replaced by the temperature difference $\Delta T$. Thus, the generalized Landau-Lifshitz-Gilbert (LLG) equation is expressed as
\begin{equation}
\label{LLG}
\begin{split}
 \frac{d\textbf{M}}{dt} = -\gamma \textbf{M}\times\textbf{H}_\mathrm{eff}
  +  \frac{\alpha_\mathrm{0}}{M_\mathrm{s}}\textbf{M}  \times & \frac{d\textbf{M}}{dt}  \\  - \frac{\gamma J_\mathrm{S}^\mathrm{STT}}{M_\mathrm{s}^2 V_\mathrm{F}}  \textbf{M}\times (\textbf{M} \times \boldsymbol{\sigma}), 
 \end{split}
 \end{equation}
where $\textbf{M}$ is the magnetization, $\gamma$  the gyromagnetic ratio, $\textbf{H}_\mathrm{eff}$  the effective magnetic field, $M_\mathrm{s}$  the saturation magnetization, $V_\mathrm{F}$  the volume of the YIG layer, and $\boldsymbol{\sigma}$   the spin polarization vector. The Gilbert damping $\alpha= \alpha_\mathrm{F}+\Delta \alpha_\mathrm{SP}$ is the sum of the intrinsic damping constant $\alpha_\mathrm{F}$ of the isolated YIG layer and $\Delta \alpha_\mathrm{SP}$ is the additional damping due to spin pumping in the adjacent Pt layer. \cite{Tserkovnyak} $J_\mathrm{S}^\mathrm{STT}$ describes the heat-induced spin torque. By introducing the injection and charge current conversion efficiency $u=  ({e}/{\hbar})({2 \pi f M_\mathrm{s} d_\mathrm{F}}/{\gamma})v$, where $v$ is the slope obtained by fitting our results (Fig.~\ref{fig:results}(b)), and by introducing an additional temperature dependent damping parameter $\Delta \alpha_\mathrm{STT}^\mathrm{SSE}$, we obtain, in analogy to Ref.~\cite{Ando_PRL_2008}, the heat-induced spin torque

\begin{equation}
\label{J_STT}
J_\mathrm{S}^\mathrm{STT} = A_\mathrm{F} v \frac{2 \pi f M_\mathrm{s} d_\mathrm{F}}{\gamma}\Delta T.
\end{equation}
The spin-current density is given by $J_\mathrm{S}= {2e}/({\hbar A_\mathrm{F}})J_\mathrm{S}^\mathrm{STT}$.

\begin{table}[t]

\centering  
\begin{tabular}{|c || c || c|} 

\hline                    
$P_\mathrm{MW}$ (dBm) &  $J_\mathrm{S}^\mathrm{STT} ( \times 10^{-11} \frac{\mathrm{Nm}}{\mathrm{^\circ C}} \Delta T)$ &  $J_\mathrm{S} (\times 10^{9} \frac{\mathrm{A}}{\mathrm{m^2} \mathrm{^\circ C}} \Delta T)$ \\ [0.5ex] 
\hline     \hline                  
+14  & 1.74 $\pm$ 0.15 & 3.70 $\pm$ 0.32  \\ 
+20 & 2.11 $\pm$ 0.18  & 4.49 $\pm$ 0.39 \\
+25 & 2.01 $\pm$ 0.08 & 4.27  $\pm$ 0.16 \\ [1ex]      
\hline 
\end{tabular}
\label{table} 
\caption{ \label{table}  Comparison of spin torque $J_\mathrm{S}^\mathrm{STT}$ and spin current density  $J_\mathrm{s}$ for different mircowave powers $P_\mathrm{MW}$.} 
\end{table}

The calculated spin torque $J_\mathrm{S}^\mathrm{STT}$ and the spin current density $J_\mathrm{S}$ are summarized for different microwave powers in Table~\ref{table}. For these calculations $M_\mathrm{S}$ is assumed to be constant since $\Delta T$ leads to variations of only about $1 \%$ which cannot explain the observed behavior. 
It should be emphasized that our calculated heat-induced spin current density per 1$\mathrm{^\circ C}$ is one to two orders of magnitude higher than those generated by the SHE for the maximal DC voltage pulses used in Ref.~\cite{MWu2011} of $U=8$ V  ($J_\mathrm{S}=10^{8} \mathrm{A/m^2}$, Pt resistance $\approx 30$ $\mathrm{\Omega}$, DC pulse length 300 ns, repetition rate 10 ms). 

Taking the large magnitude of the observed heat-induced STT (linewidth change about 6$\%$) for the comparably small temperature difference across the actual YIG/Pt interface (less than 0.1$^\circ$C) into account, we might consider influences on $\Delta H$ by other effects:
(1) The change in electric resistance of the Pt layer due to the applied temperature difference (less than approximately $0.1\%$) cannot be the origin of the observed linewidth change.  (2) In the present experiment, phonons penetrating the entire sample stack (including the 500 $\mu$m thick GGG layer) are the main heat carriers. \cite{Uchida_long, Uchida_JAP} Consequently, they are the major cause for the thermally-induced spin current.

In conclusion, the heat-induced damping modification in YIG/Pt hetero-structures has been shown. 
The modulation of the relaxation coefficient has been demonstrated by spin pumping as well as by microwave techniques. Both techniques qualitatively show the same behavior. 
Besides that, our findings demonstrate, that every spin pumping experiment, in which the coherent magnetization precession is driven by a microwave source, is accompanied by heating. We have introduced a method to identify spin pumping from coherent magnons and SSE contribution from incoherent magnons to the ISHE voltage, resulting in an increase of the ISHE voltage of around 0.1\%. The spin transfer due to the temperature difference across the YIG/Pt interface has been estimated and has been compared to previous works that use SHE-generated spin currents. It turns out, that the heat-induced spin current density $J_\mathrm{S}$ per 1$\mathrm{^\circ C}$ is of the order $10^{9} {\mathrm{A}}/{\mathrm{m^2}}$.

We thank G.~A.~Melkov for valuable discussions and R.~Neb for the platinum film deposition. 
Financial support by the Deutsche Forschungsgemeinschaft within the projects SE 1771/4-1 (Priority Program 1538 ÒSpin Caloric TransportÓ) and CH 1037/1-1 is gratefully acknowledged.

\end{document}